\documentstyle[12pt]{article}
\begin{document}
\title{Particle scattering in nonassociative \\
        quantum field theory}
\author{Dzhunushaliev V.D.\thanks{e-mail: dzhun@freenet.bishkek.su}}

\date{}

\maketitle
\begin{center}
Theoretical physics department, the Kyrgyz State National
University, 720024, Bishkek, Kyrgyzstan
\end{center}

\begin{abstract}
{
A model of quantum field theory in which the field operators form
a nonassociative algebra is proposed. In such a case, the n-point
Green's functions become functionally independent of each other.
It is shown that particle interaction in such a theory can
be realized by nonlocal virtual objects.
}
\end{abstract}

\centerline{PACS N 03.70.+k}

At the present time, attempts are being made to introduce in some
form or other nonassociative classical fields in field
theory \cite{quasi}. In this paper, we propose a modification of the rules
for quantization of field operators in which the algebra of field
operators becomes nonassociative. Such a modification has the consequence
that there appear in the quantum theory Green's
functions that are functionally independent of each other. One of the
interesting consequences of such a quantum theory is that
in it there arise Feynman diagrams for particle scattering that cannot
be interpreted as the result of the exchange of virtual quanta
as carriers of the interaction.

\par
In standard quantum field theory, the interaction of elementary particles
takes place through the exchange of virtual
particles. The main building bricks in the construction of the theory
of interacting quantum fields are the propagators of the
corresponding particles and their interaction vertices. In string theory,
particle-particle scattering is implemented by means
of nonlocal objects - strings. In this connection, the following question
arises: Is it possible to modify the rules of field
quantization in such a way that the elementary building brick is not
a 2-point Green's function but, for example, a 4-point
Green's function (as occurs in the Veneziano dual model and in string
theory \cite{green})? If such a modification is possible, then it
may be supposed that the interaction between elementary particles
(described by the 4-point Green's function) in such a modified
field theory will be implemented by nonlocal objects.

\par
To obtain a 4-point Green's function (with postulation of certain
quantization rules), it is necessary to relate in some
nontrivial manner the quantum field operators at four space time points.
In standard field theory, such a procedure, applied to
the product of two field operators, leads to the appearance
of the 2-point Green's function (propagator), as a consequence of
which the order of succession of the operators in the product of
field operators mentioned above becomes important.
\par
It is clear that in noncommutative quantum field theory it is impossible
to define a 4-point Green's function that is not some
polylinear combination of propagators of the corresponding particles
interacting at the vertices. Thus, a significant modification
of the quantization rules is necessary. It turns out that to define
such a Green's function it is necessary to require that the algebra
of the quantized field become nonassociative. Then a nontrivial
4-point Green's function appears when the time-ordered product
of four field operators is defined as follows: When any field
operator is pulled through, in its place it is necessary to redefine
the order of the bracket in the product of field operators.
Each such redefinition leads to the appearance of an associator, just
as in noncommutative algebra displacement of an operator to the left
or right gives rise to the appearance of the commutator.
It is obvious that the number of operators needed to define
the associator cannot be less than three. In such a nonassociative
quantum field theory, we can obtain n-point Green's functions
that are independent of each other.
\par
We now turn to a more detailed analysis of nonassociative quantum field
theory. We denote the product of $n$ field operators
in which the brackets are arranged in accordance with some rule
$P$ by $M_n(P)$. For example,
\begin{equation}
M_3(P) =\Bigl ((\hat \varphi _x \hat \varphi _y )\hat \varphi _z\Bigl ),
\label{1}
\end{equation}
where $\hat\varphi _x = \hat\varphi (x)$
is the field operator at the point $x$;
in this case, rule $P$ indicates that all opening brackets are in the
extreme left hand position.
\par
We shall call this ordering of the brackets for any number of operators,
i.e. , with all opening brackets in the extreme left
hand position, the normal position of the brackets and denote it by
means of the colon:
\begin{equation}
: M_n(P) := \Biggl (\biggl (\cdots \Bigl (\hat \varphi _1(
\hat\varphi _2 \cdots \biggl )\hat\varphi _n\Biggl ),
\label{2}
\end{equation}
here we have on the left $n$ opening brackets in succession.
\par
We consider the nonassociative quantum field theory that satisfies
the following axiom.
\par
I. Two monomials $M_n(P_1)$ and $M_n(P_2)$ that differ only in the
rule of placing of the brackets differ from each other by a
numerical function $Ass(x_1,x_2,\ldots ,x_n; P_1, P_2)$,
which we call the n-point associator:
\begin{equation}
M_n(P_1) - M_n(P_2) = Ass(x_1,x_2,\ldots ,x_n;P_1,P_2).
\label{3}
\end{equation}
\par
2. The action of the product of two monomials on the quantum state
vector $|v>$ is defined as follows:
\begin{eqnarray}
\Bigl ( M_n(P_1) M_k(P_2) \Bigl ) |v> & = &
M_n(P_1) \Bigl (M_k(P_2)|v> \Bigl ),
\label{41}\\
<v| \Bigl ( M_n(P_1) M_k(P_2) \Bigl ) & = &
\Bigl ( <v| M_n(P_1) \Bigl ) M_k(P_2).
\label{42}
\end{eqnarray}
\par
3. We have the usual commutation rules
\begin{equation}
\left [ \hat \varphi _x , \hat \varphi _y \right ] = G(x,y),
\label{5}
\end{equation}
where $G(x, y)$ is the 2-point Green's function (propagator).
\par
For more transparent penetration to the essence of such a nonassociative
quantum field theory, we consider the following
simplified model, in which the action of any field operator on the
vacuum state annihilates it
\begin{equation}
\hat\varphi _x |vac> = <vac|\hat\varphi _x = 0.
\label{6}
\end{equation}
We now consider the 3-point assocator
\begin{equation}
(\hat\varphi _x\hat\varphi _y)\hat\varphi _z =
\hat\varphi _x(\hat\varphi _y\hat\varphi _z) +
Ass(x,y,z).
\label{7}
\end{equation}
Using the rule introduced above for the action of the field operators
on the vacuum state, we readily deduce that $Ass(x,y,z)=0$.
We now define two 4-point assocator:
\begin{eqnarray}
\Bigl ((\hat\varphi _x\hat\varphi _y)\hat\varphi _z\Bigl )\hat\varphi _u
& = & (\hat\varphi _x\hat\varphi _y)(\hat\varphi _z\hat\varphi _u) +
Ass_1(x,y|z,u),
\label{81}\\
\hat\varphi _x\Bigl (\hat\varphi _y(\hat\varphi _z\hat\varphi _u)\Bigl )
& = & (\hat\varphi _x\hat\varphi _y)(\hat\varphi _z\hat\varphi _u) +
Ass_2(x,y|z,u),
\label{82}
\end{eqnarray}
Acting on each of Eqs.(\ref{81}-\ref{82}) from the right and left with
the vacuum state and using the condition (\ref{6}), we can express the 4-point
assocators $Ass_{1,2}$ in terms of the vacuum expecation value
of the monomial
$(\hat\varphi _x\hat\varphi _y)(\hat\varphi _z\hat\varphi _u)$:
\begin{equation}
Ass_{1,2}(x,y|z,u) = Ass(x,y|z,u) = -<vac|
(\hat\varphi _x\hat\varphi _y)(\hat\varphi _z\hat\varphi _u)|vac>.
\label{9}
\end{equation}
We introduce some properties of the 4-point associator in the case
when the points $(x^\mu$ and $y^{\mu})$ are pairwise
spacelike:
$x^{\mu}\sim y^{\mu},z^{\mu}\sim u^{\mu}$
($x^{\mu}, y^{\mu}, z^{\mu}$ and $u^{\mu}$
are 4-dimensional vectors). Using the commutation properties
of the operators,
and also the properties of the vacuum state, we can esablish
the following symmetry properties of the 4-point assocator:
\begin{equation}
Ass(x,y|z,u) = Ass(y,x|z,u) = Ass(x,y|u,z).
\label{10}
\end{equation}
We now define the Green's function as follows:
\begin{equation}
T(M_n(P)) = M_n(P) + G(x_1,x_2,\ldots ,x_n;P),
\label{11}
\end{equation}
where the symbol $T$ denotes the ordinary time ordering of the operators
in the monomial $M_n(P)$, and also normal ordering of
the brackets in the monomial $M_n(P)$. It is obvious that the
n-point Green's function can be expressed in terms of a polylinear
combination of $n$-point assocators $(m\le n)$ and propagators.
\par
We now consider the interaction of two quantized fields.
The operators of one field $\hat q(x)$ form an ordinary noncommuative
algebra, but the operator of the other field $\hat\varphi (x)$
form the nonassociative algebra proposed above. We define the simplest 
model Lagrangian of their interaction as follows:
\begin{equation}
L_{int}(x) = i \hat q_x \gamma ^{\mu} \hat A_{x\mu} \hat q_x,
\label{12}
\end{equation}
Despite the apparent triviality of (\ref{12}), in this model theory
there is a very nontrivial interaction of these fields due to the
nonassocative nature of the quantized field $\hat\varphi (x)$.
\par
As an example, we reduce to normal form the $T$ product
of the four operators of the interaction Lagrangian:
\begin{equation}
S_4 \left ( x,y|z,u \right ) =
T \left (
J^{\alpha}(x) J^{\beta}(y) J^{\gamma}(z) J^{\delta}(u) \right )
T\left ( : A_{\alpha x} A_{\beta y} A_{\gamma z} A_{\delta u}
:\right ).
\label{13}
\end{equation}
Suppose the field operators $\hat q$ and $\hat\varphi$
commute with each other;
then all the field operators $\hat q$ can be collected together in one
bracket, for example, on the left:
\begin{eqnarray}
S_4(x,y|z,u) & = &
T\Biggl [:\hat q_x\hat q_y\hat q_z\hat q_u::
\Biggl (\biggl ((
\hat\varphi _x\hat\varphi _y)\hat\varphi _z\biggl )
\hat\varphi _u\Biggl ) \Biggl ] =
\nonumber \\
& &
T[:\hat q_x\hat q_y\hat q_z\hat q_u:]
T\Biggl [\Biggl (\biggl ((
\hat\varphi _x\hat\varphi _y)\hat\varphi _z\biggl )
\hat\varphi _u\Biggl ) \Biggl ].
\label{14}
\end{eqnarray}
We shall be interested in only the term
\begin{equation}
D_4(x,y|z,u) = :\hat q_x\hat q_y\hat q_z\hat q_u:
G(x,y|z,u),
\label{15}
\end{equation}
where in accordance with (\ref{11})
\begin{equation}
G(x,y|z,u) = T\Biggl [\Biggl (\biggl ((\hat\varphi _x\hat\varphi _y)
\hat\varphi _z\biggl )\hat\varphi _u \Biggl )\Biggl ].
\label{16}
\end{equation}
This diagram describes the scattering of four quanta of the $\hat q$
field realized by means of the 4-point Green's function of the
nonassocative quantized field $\hat\varphi$, whose operator algebra
is nonassocative and noncommutative.
In contrast to an ordinary Feynman diagram, in this case one cannot
speak of the exchange of quanta of the $\hat\varphi$ field,
since the 4-point Green's function $G(x,y|z,u)$ is not any
combination of propagators $G(x,y)$ of the field $\hat\varphi$.
\par
We show that the scattering amplitude (\ref{15}) has certain features
in common with the Veneziano amplitude. Because from
the beginning we took nonassociative field operators into our theory,
the amplitude (\ref{15}) cannot, in general, be expressed in
terms of the field propagators $G(x,y)$. The Veneziano amplitude $A(s,t)$,
\begin{equation}
A(s,t) = - \sum _{k=0}^{\infty} \frac{[\alpha (s) + 1]
[\alpha (s) + 2]\ldots [\alpha (s) + k]}{k!}
\frac{1}{\alpha (t) - k},
\label{17}
\end{equation}
($\alpha (t) = a' + \alpha(0)$ is a "Regge trajectory") has a similar
property, but in a somewhat weakened form. It is an infinite sum, each
term of which corresponds to exchange of some particle. But precisely
because the sum is infinite, the result after summation
differs from any finite sum that approximates it:
\begin{equation}
A(s,t) = - \sum _{k=0}^{n} \frac{[\alpha (s) + 1]
[\alpha (s) + 2]\ldots [\alpha (s) + k]}{k!}
\frac{1}{\alpha (t) - k},
\label{18}
\end{equation}
This difference takes the form that although each finite sum $A_n(s,t)$
is an entire junction of $s$ nevertheless the Veneziano
amplitude $A(s,t)$ has a pole with respect to $s$, which ensures
its crossing symmetry.
\par
We now consider the following property of the amplitude (\ref{15}).
We recall that we consider two spacelike pairs $x^{\mu}\sim y^{\mu}$ and
$z^{\mu}\sim u^{\mu}$. The expression (\ref{15}) indicates that there
is an interaction between the pairs $\hat q_x\hat q_y$ and
$\hat q_z\hat q_u$, which
must be implemented by some nonlocal virtual object.
If we take each instant of time $\tau$ between
$t^1=x_0=y_0$ and $t_2=z^0=u^0$ $(x^0, y^0, z^0$ and $u^0$
are the time components of the 4-vectors
$x^{\mu}, y^{\mu}, z^{\mu},$ and $u^{\mu}$), then for each $\tau$
the Green's function $G(x,y|z,u)$ is a nonlocal virtual
object that implements interaction between the pairs
$\hat q_y\hat q_y$ and $\hat q_z\hat q_u$. This observation indicates
that this model in spirit is similar to the string model,
in which the interaction between the pairs mentioned above is
implemented by a virtual string, however, there is the important
difference that in the nonassociative quantum field theory that we
consider the nonlocal object $G(x,y|z,u)_{t=\tau}$ is not one dimensional.
\par
Thus, we see that in nonassociative quantum field theory interaction
between particles can be realized by nonlocal virtual
object. In this sense, such a quantum field theory is to some
degree similar to string theory, in which particle scattering can
be described by nonlocal extended objects such as strings.

\end{document}